\documentclass[11pt]{iopart}
             
\usepackage{graphicx}
\usepackage{hyperref} 

\begin{document}

\title[``\textit{Ultima Ratio}'': Simulating wide-range X-ray scattering and diffraction]{``\textit{Ultima Ratio}'': Simulating wide-range X-ray scattering and diffraction}

\author{Brian R. Pauw, Sofya Laskina, Aakash Naik, Glen J. Smales, Janine George, Ingo Bre{\ss}ler, Philipp Benner} 
\address{Bundesanstalt f\"ur Materialforschung und -Pr\"ufung, Unter den Eichen 87, 12205 Berlin, Germany}


\ead{brian.pauw@bam.de}
\vspace{10pt}
\begin{indented}
\item[]March 2023
\end{indented}

\begin{abstract}
We demonstrate a strategy for simulating wide-range X-ray scattering patterns, which spans the small- and wide scattering angles as well as the scattering angles typically used for Pair Distribution Function (PDF) analysis. Such simulated patterns can be used to test holistic analysis models, and, since the diffraction intensity is on the same scale as the scattering intensity, may offer a novel pathway for determining the degree of crystallinity. 

The ``Ultima Ratio'' strategy is demonstrated on a 64-nm Metal Organic Framework (MOF) particle, calculated from $Q<0.01$\,$\mathrm{nm}^{-1}$ up to $Q\approx150$\,$\mathrm{nm}^{-1}$, with a resolution of 0.16\,\AA. The computations exploit a modified 3D Fast Fourier Transform (3D-FFT), whose modifications enable the transformations of matrices at least up to $8000^3$ voxels in size. Multiple of these modified 3D-FFTs are combined to improve the low-$Q$ behaviour.  
The resulting curve is compared to a wide-range scattering pattern measured on a polydisperse MOF powder. 

While computationally intensive, the approach is expected to be useful for simulating scattering from a wide range of realistic, complex structures, from (poly-)crystalline particles to hierarchical, multicomponent structures such as viruses and catalysts.

\end{abstract}

\section{Introduction}

Functional materials derive their properties from a complex interplay of hierarchical structures, from the atomic scale up to the micrometer scale and beyond. Such hierarchical structures sometimes lend themselves to investigation with modern X-ray and neutron scattering methodologies. These now measure scattering over several decades of scattering vector (Q) by default, thus carrying information on several decades in structural length scales. This wide measurement range invites the construction of analysis models that can interpret the entirety of the expanded view. The output of such analysis models should be tested against - or compared with simulated data from fully understood structures, but in practice such simulations prove challenging to construct. 

Fourier Transformation (FT) of a simulated electron density is a close mathematical analog to the physical scattering process, and can be used to compute X-ray and neutron scattering patterns accurately \protect{\footnote{To simulate neutron scattering, the nuclear density (and magnetic structure if applicable) rather than the electron density is used as the input.}}. These simulated patterns include both scattering as well as diffraction phenomena, as diffraction is, in essence, scattering from regular structures \*), and can extend well into the Pair Distribution Function (PDF) range. One major advantage of FT is that the simulation can be made as close to reality as required. Its unrestricted nature allows for hierarchical electron density maps to be simulated, and even instrumental effects are implementable, such as limited coherence, smearing effects and wavelength dispersion. 

Before it can be used, challenges with this approach remain to be addressed. In particular the explosive scaling of computational resources upon modest increases of simulated box resolution appears at first glance to be insurmountable. For example, while an electron density (ED) box of 512$^3$ equidistant points can be Fourier Transformed within a minute on normal hardware, doubling this resolution for a nearly insignificant amount more Q will exhaust most computer systems' memory. Conversely, increasing the ED sampling point distance leads to a loss of important details, resulting in reduced predictive qualities at high Q. While it is possible to catapult more hardware at this problem \cite{nVidia-2022}, such solutions are outside of the budget of most laboratories. Other solutions are therefore needed. 

This short communication explores two approaches that, together, can help the small laboratory simulate using high-resolution Fourier transforms. Firstly, we demonstrate the ability to generate a high-resolution electron density (ED) map of a crystalline nanoparticle in a slice-by-slice manner by tiling and masking an \textit{ab-initio} electron density simulation of a crystalline unit cell. The thus obtained ED can be Fourier Transformed in a slice-by-slice manner to obtain high-resolution simulated scattering information. This partially circumvents the problem of memory restrictions (restricting instead to the available disk space). Secondly, we obviate the need for real-space zero padding by merging simulations: one of the primary particle (ED-1), combined with simulations of smaller, ``zoomed out'' densities of the lone particle from ED-1 in ED-2 and ED-3. Finally, we combine these approaches into a workable strategy called "Ultima Ratio"\protect{\footnote{After ``\textit{Ultima Ratio Regum}'', historically stamped on cannons commissioned by King Louis XIV, and literally translates as ``the last argument of the King''. Like its namesake, the simulation software here only becomes relevant when all other assumptions fail, and should be considered a rather blunt instrument to solve this particular problem.}}, for simulating more complex hierarchical structures. This is demonstrated in a software demonstrator (Jupyter notebook), which, together with the necessary supplemental files are available under a CC-BY-4.0 license from Zenodo \cite{Pauw-2023}. With the pathway proven in this short communication, a fully featured solution becomes attainable. 

\section{Methodology}

\subsection{General Workflow}

\begin{figure}[ht]
\begin{center}
\includegraphics[width=0.7\textwidth]{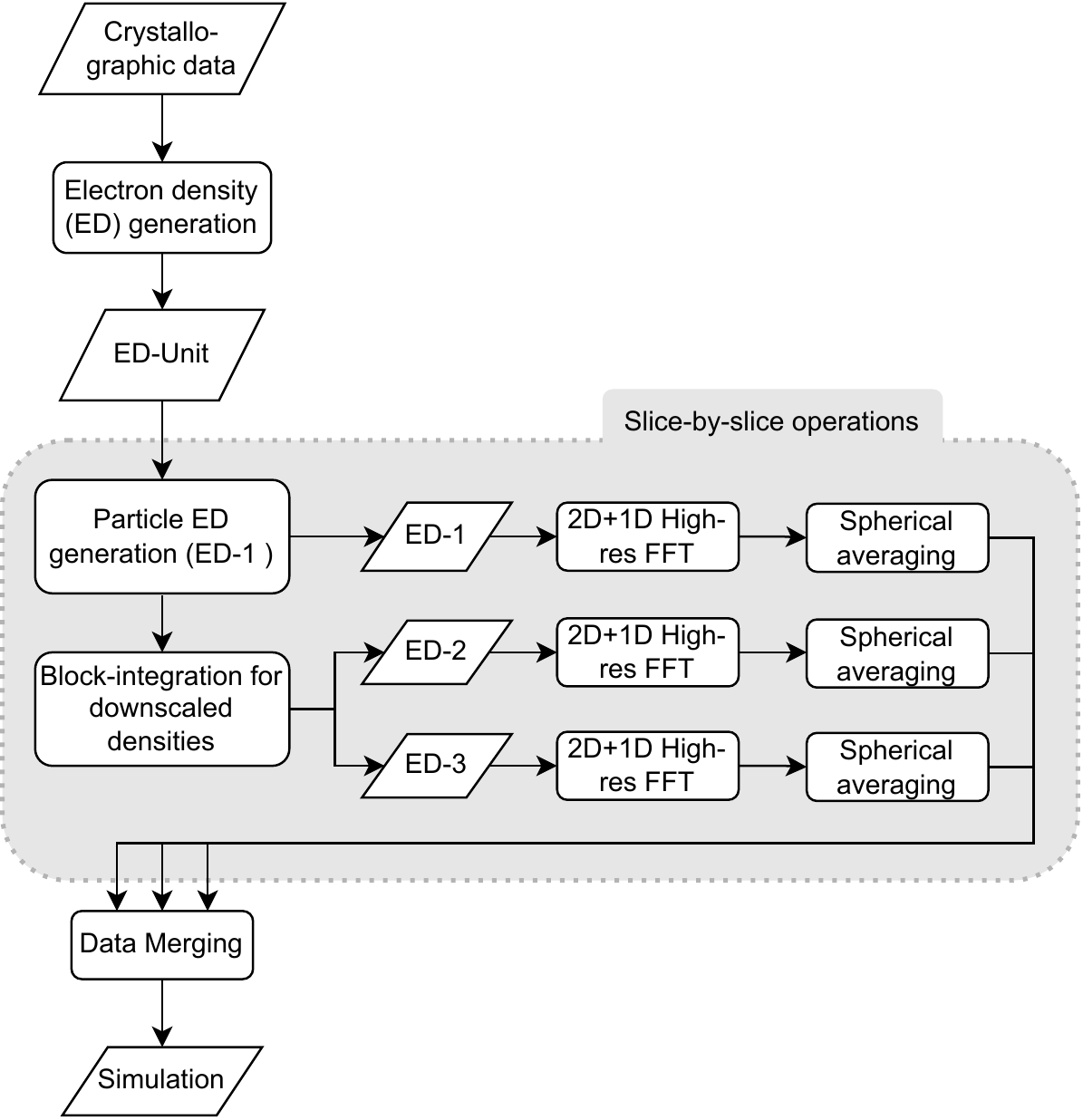}
\caption{\emph{Ultima Ratio}-workflow for obtaining wide-range X-ray scattering simulations. The combination of multiple scales and separated 2D+1D FFT steps (to achieve a 3D FFT without the memory requirements) enable the simulation of scattering from realistic densities and (hierarchical) structures. } 
\label{fg:workflow}
\end{center}
\end{figure}

To demonstrate the overall strategy, a wide-range scattering pattern is simulated for a representative metal-organic framework (MOF) particle of the Zeolite Imidazole Framework (ZIF-8) chemistry \cite{Karagiaridi-2012}. 
The general workflow used to achieve this is schematically visualized in Figure \ref{fg:workflow}, and detailed in the following paragraphs. In short, these steps are:
\begin{enumerate}
\item{} An electron density map for the unit cell (ED-Unit) is generated. 
\item{} ED-Unit is then tiled within a particle boundary to generate a large, high-resolution ED map of the particle (ED-1). 
\item{} ED-1 is block-integrated stepwise to obtain lower-resolution ED-2 and ED-3 maps of the ``zoomed out'' structure.
\item{} The particle at different zoom levels, ED-1, ED-2 and ED-3 are zero-padded and 3D-Fourier Transformed in slices using the separated 3D-FFT.
\item{} A spherical averaging step is done on each obtained FFT, to obtain a 1D curve of intensity $I$ vs. scattering vector $Q$.
\item{} A merging step is done to combine the curves into a single dataset. 
\end{enumerate}
With all but the first and last step performed on a slice-by-slice basis with intermediate hard disk storage of the full matrices stored in the Hierarchical Data Format (HDF5), which support lazy loading of arrays and multiple simultaneous read operations \cite{Folk-2010}.

\subsection{Density functional theory computational details}

An all-electron charge density map for a MOF conventional unit cell (ED-Unit) of $1.683^3$\,nm$^3$ is computed using density functional theory. The initial structure of ZIF-8 is retrieved from Crystallography Open Database (COD ID: 4118891)\cite{Karagiaridi-2012, Gravzulis-2012, Downs-2003}. The structural model has partial occupancies for C and H atoms. For the sake of simplicity, we only retain the sites with higher partial occupancies (0.630) for C and H, using VESTA\cite{Momma-2011}. 

VASP 5.4.4 \cite{Kresse-1993, Kresse-1996a, Kresse-1999} is used with the projector augmented wave (PAW) \cite{Blochl-1994} pseudopotentials (PP) for the density functional theory computations. The exchange-correlation energy is treated with the Perdew–Burke–Ernzerhof (PBE)\cite{Perdew-1996} functional. Structural optimization without volume relaxation is performed twice. For this optimization runs, the plane wave cutoff energy is set to 520 eV, the k-point grid to $1\times1\times1$, the electronic energy convergence criterion to $10^{-6}$ eV, and the ionic convergence criterion to $10^{-5}$ eV.

Next, the cutoff energy and $\textbf{k}$-point grid convergence tests are performed on the relaxed structure. The reliability of all-electron charge density files is validated by performing Bader charge analysis\cite{Bader-1990, Tang-2009, Sanville-2007}. The all-electron charge density is considered reliable if the total electron count (including core electrons) based on Bader partitioning equals the total number of electrons in the structure and the Bader charges obtained for each atom converge. Using a cutoff energy of 500 eV,  a $\Gamma$-centered $\textbf{k}$-point grid of $2\times2\times2$ and an FFT grid equal to 630 points during VASP static runs turned out to be the optimal setting to meet the criterion. The results of these convergence tests are provided in the following data repository \cite{Pauw-2023}. For ED-Unit, the error in total electrons is ~ 0.023\,\% ($1200.227/1200$).

\subsection{Electron density generation}

The ED-Unit generated above has dimensions of $630^3$ voxels, and are provided as 32-bit floats. Before using this in our simulations, we reduce the resolution \emph{via} the block integration functionality of the scikit-image library (\verb|skimage.measure.block_reduce|) [https://doi.org/10.7717/peerj.453]. The reduction factor is chosen such that 1) the VASP unit cell is divisible by it, and that 2) the maximum scattering vector $Q$ remains well within the simulated range. For this simulation, the reduction factor is set to 6, generating a reduced ED-Unit of 105$^3$ voxels. Combined with the unit cell lattice dimensions above, this results in a sampling step length after reduction of 0.16 \AA ngstr\"om. 

This reduced ED-Unit is then interpolated over a rotated 2D planar grid (the rotation is performed to demonstrate the practical possibility of generating polycrystalline materials with random crystallite orientations), with ED-Unit repeated outside of its boundaries. This exploits the "\verb|map_coordinates|" method of scipy in "\verb|grid-wrap|" mode (\verb|scipy.ndimage.map_coordinates|). Each 2D planar grid is then multiplied with a circular mask dependent on the coordinate of the plane, with the mask defining the external particle boundary. This plane is shifted by one voxel-length at a time along its normal, so that a 3D electron density (ED-1) is generated in a slice-by-slice manner, thus describing a spherical particle filled with a lattice of unit cells. 

For improved low-Q simulations, additional (zero-)padding is required around the object of interest in the simulation box. This is effected through "zooming out" from the object of interest, thus creating ED-2 and ED-3, achieved through slice-by-slice block integration and zero padding of ED-1 with factors of 16 and 125, respectively. As high resolution is no longer of the utmost interest at this point, the box voxel resolution of ED-2 and ED-3 is simultaneously reduced to a faster, smaller box. Both integration as well as resolution reduction are achieved in a single block integration step, but two orthogonal integration steps are necessary (the first on the 2D images, the second using a 1D kernel in the remaining direction). The resulting electron densities are stored in HDF5 files as 3D datasets using 32-bit floats. ED-1, ED-2 and ED-3 are shown in Figure \ref{fg:EDScales}.

As for the dimensions of ED-1, ED-2 and ED-3, it should be understood that the FFT method implicitly assumes periodic boundary conditions (PBCs). These PBCs may be exploited in simulations of structures that contain one or more repeating elements. However, if one is to prevent interference from its periodic-boundary neighbour, as is the case for these isolated MOF particles, the object of interest should not occupy more than half of a zero-padded box. Thus, while we want to compute an FFT of our 64-nm particle in a box containing $8000^3$ voxels, our particle will only occupy $4000^3$ voxels, and so the remaining density does not need to be written, saving a factor of 8 in disk i/o. The remaining voxels are added during the FFT procedure. The density-of-interest of ED-2 and ED-3 is therefore proportionally smaller, and their sizes of $250^3$ and $32^3$ will fit comfortably in FFT boxes containing $2000^3$ voxels in total.

\subsection{Fourier Transformation}

The 3D Fast Fourier Transform (FFT) is carried out by splitting the Fourier transform into $n$ 2D FFTs, followed by $n^2$ 1D FFTs in the orthogonal direction. The 3D FFT of an electron density is defined as:
\begin{equation}
\label{eq:FT_3D}
    F(u,v,w) = \frac{1}{MNL}\sum_{x=0}^{M-1}\sum_{y=0}^{N-1}\sum_{z=0}^{L-1}\rho_{xyz} \cdot e^{-i2 \pi (\frac{ ux}{M}+\frac{vy}{N} + \frac{wz}{L})},
\end{equation}

where the electron density $\rho_{xyz}$ is now given by a sampling point at $(x,y, z)$ coordinates in 3D space and $u$, $v$ and $w$ describe each of the 3 domains in space the signal is measured on.

This can be separated into $n$ 2D and $n^2$ orthogonal 1D components, by separating one exponential term to get us our working method:
\begin{equation}
\label{eq:FT_2D1D}
    F(u,v,w) = \frac{1}{M}  \sum_{x=0}^{M-1} e^{-i2\pi\frac{ u x }{M}} \left(\frac{1}{NL}  \sum_{y=0}^{N-1} \sum_{z=0}^{L-1} \rho_{xyz}e^{-i2 \pi(\frac{ v y}{N}+\frac{ w z}{L})}\right)
\end{equation}
These parts can be computed separately with an intermediate disk storage between the 2D and the 1D FFTs, reducing the memory requirements from $\mathcal{O}(n^3)$ to $\mathcal{O}(n^2)$ (although the disk storage requirement for the final 3D FFT remains $\mathcal{O}(n^3)$. An additional advantage to this separation is that for the initial 2D FFT, only the slices with nonzero density need to be computed, which saves a factor of two in space and time for the intermediate storage and calculation, respectively. 

For the FFTs, the PyTorch library is used (\verb|torch.fft|), so that the tasks may be offloaded to a cuFFT-capable GPU when such hardware is available \cite{Paszke-2019}. Before storage in the intermediate HDF5 structure, the 2D FFTs are fft-shifted so that the lowest frequency is in the center. After the 2D FFT slices are stored, a series of 1D FFTs are performed in the orthogonal direction and the output stored in a new HDF5 file. The intermediate FFT is stored as 128-bit complex numbers to avoid loss of precision, the absolute squared amplitude of the final FFT (the intensity) is stored as 32-bit float. As mentioned, to save computational power, the first 2D FFT is performed only on slices with density, with the second FFT padding the uncalculated empty slices.

\subsection{Q-calculation and spherical averaging}

If there was an interest in the anisotropic properties of this sample, 2D data reflecting a practical experiment could be extracted by modeling a spherical surface (i.e. the Ewald sphere) with a shell thickness defined by the wavelength distribution, and extracting the intensity along that surface. Given the isotropic nature of the experimental MOF data, however, 2D intensity data is not needed. Therefore, spherical averaging is done over the entire matrix on a slice-by-slice basis. For each slice, a $Q$ map is computed based on the length information of the sides of the simulation box, the number of voxels per side, and the distance of the slice from the center.

Both the intensity slice and the Q map are fed into our dataMerge software for binning (c.f. Appendix D of \cite{Smales-2021}). This binning procedure simultaneously averages the intensity and determines the standard error on the mean (and a few more statistics), within each of the logarithmically-spaced bin edges. The core of this software has been rewritten in compiled Python (cython), resulting in a significant speed boost.

\begin{figure}[ht]
\includegraphics[width=0.99\textwidth]{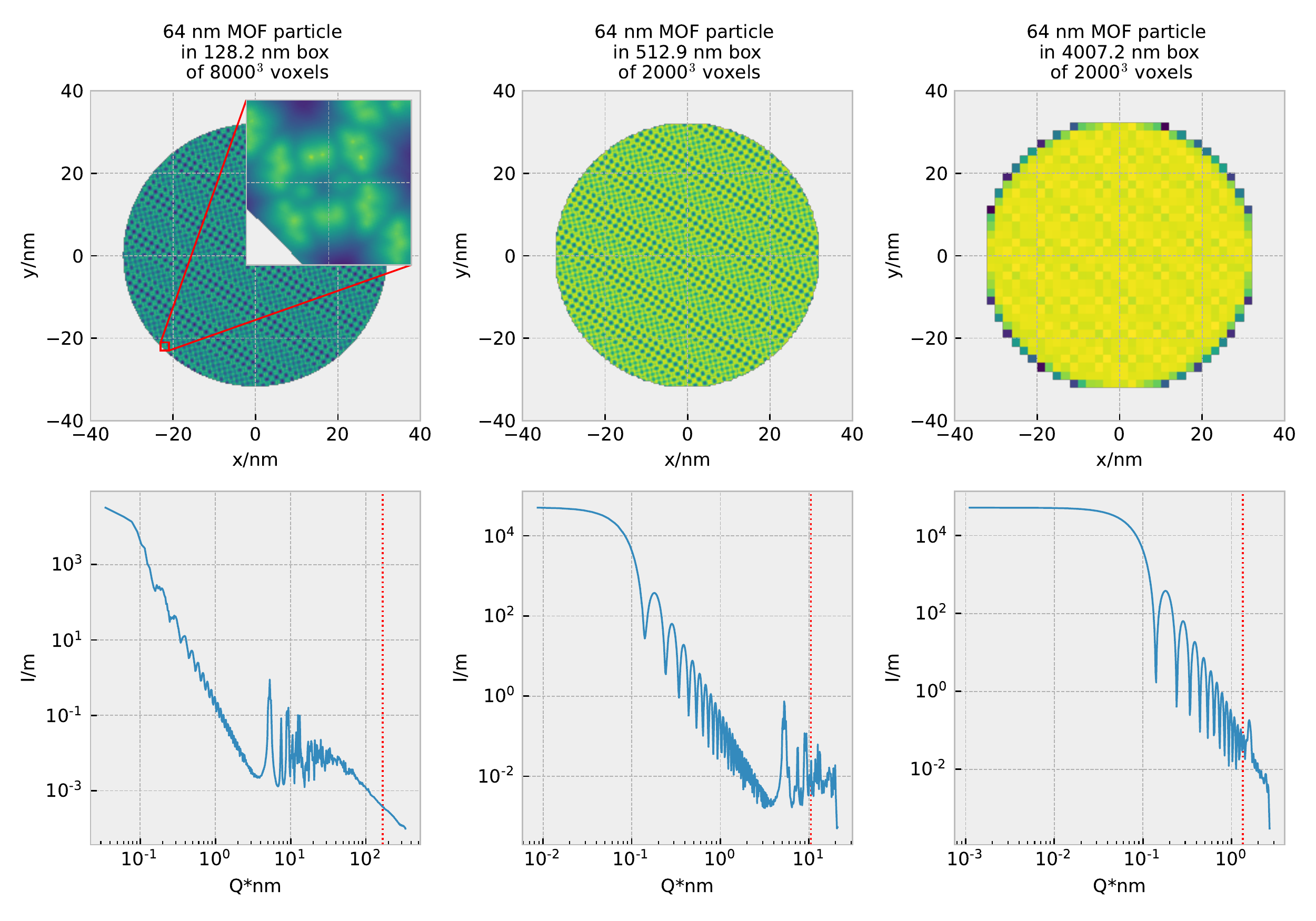}
\caption{Top row: the central slice through the electron density maps ED-1 (left), ED-2 (center) and ED-3 (right). Only the interesting portion is shown, the padded zeros are omitted. For ED-1, the inset shows a 2 by 2\,nm zoom on the electron density map. In this color map, blue is low density, yellow is high density.  Bottom row: the scattering patterns computed for these structures. The red dashed line shows the limit beyond which artefacts can be identified. This appears at $Q>Q_{max}/2$.} 
\label{fg:EDScales}
\end{figure}

The thus obtained per-slice averaged intensity then has to be combined in a separate averaging step. This is again done by dataMerge, propagating the previously determined uncertainty estimates. Thus, a single curve is obtained for each high-resolution simulation (as shown in the bottom half of Figure \ref{fg:EDScales}). Finally, a third dataMerge step combines the curves from the multiple scale simulations. It should be noted that while in the past we additionally convoluted our scattering pattern with a voxel-sized sinc function for improved high-Q behaviour (c.f. \cite{Pauw-2010}), here we have omitted this step for its negligible impact with these tiny voxels. It may be considered for inclusion again when precision is needed close to the upper $Q$-boundary.

\subsection{Details on computational hardware and a note on performance}

The simulations are run on a 2-CPU Hewlett-Packard Z8 G4 workstation running Ubuntu Linux 22.04.2 LTS. This workstation contains two Intel Xeon Silver 4108 CPUs running at nominally 1.8 GHz. 384 GB internal RAM is available, in addition to 20 TB scratch space spread across three spinning disks set up in a RAID-0 stripe configuration. Computations are done in JupyterLab with a Python 3.10 kernel. The above calculation requires about 6 TB of hard disk space.

While the generation of ED-1, ED-2 and ED-3 is sufficiently fast, the calculation of the FFT of ED-1 at $8000^3$ takes three days (compared to ED-2 and ED-3, which take about 20 minutes each). This long processing time is largely due to an i/o bottleneck on the spinning disks, as a $4000^3$ FFT only takes approximately four hours. Up to, and including $4000^3$ voxels, the CPU cores remain fully occupied, but are idling significantly at $8000^3$ while i/o transfer is pegged. It is therefore recommended to attach faster SSD storage instead if budget allows and large FFTs are needed. Some of these issues may be addressed by improved code as well.

\subsection{Synthesis and Characterisation}

ZIF-8 (Zinc Imidazole Framework-8) was synthesised from two stock solutions, the first consisting of zinc nitrate hexahydrate in methanol (MeOH), and the second consisting of 2-Methylimidazole (2-MeIm) in MeOH. 10 ml of each stock solution was injected into a falcon tube, at a rate of 15 ml/min, and stirred at 200 rpm for 21.5(5) minutes at an ambient laboratory temperature of 28$^{\circ}$C. This resulted in a final synthesis of Zn: 2-MeIm: MeOH molar ratio of 1: 7.5: 290. After the allowed synthesis time, the reaction mixture was centrifuged at 6000 rpm for 20 minutes and subsequently dried at 60$^{\circ}$C for 22 hours. The as-synthesised powder was then chracterised using SAXS/WAXS.

SAXS/WAXS measurements were performed using the MOUSE instrument where a microfocus monochromatized X-ray source ($\lambda_\textrm{Cu} = 0.154$ nm) was used \cite{Smales-2021}. Data was collected on an in-vacuum Dectris Eiger R 1M detector placed at multiple distances between from ca. 57 - 2507 mm from the sample. The powder sample was held in the beam between two pieces of scotch magic tape\texttrademark. The resulting data was then processed using the DAWN software package, following a standardized data correction procedures \cite{Basham-2015, Filik-2017, Pauw-2017}. After correction, the data from the different distances were combined using a procedure that merges the datasets in such a way as to optimize the uncertainties in the overlapping regions, producing a single continuous curve \cite{Pauw-2022}.

\section{Results and Discussion}

\begin{figure}[ht]
\includegraphics[width=0.99\textwidth]{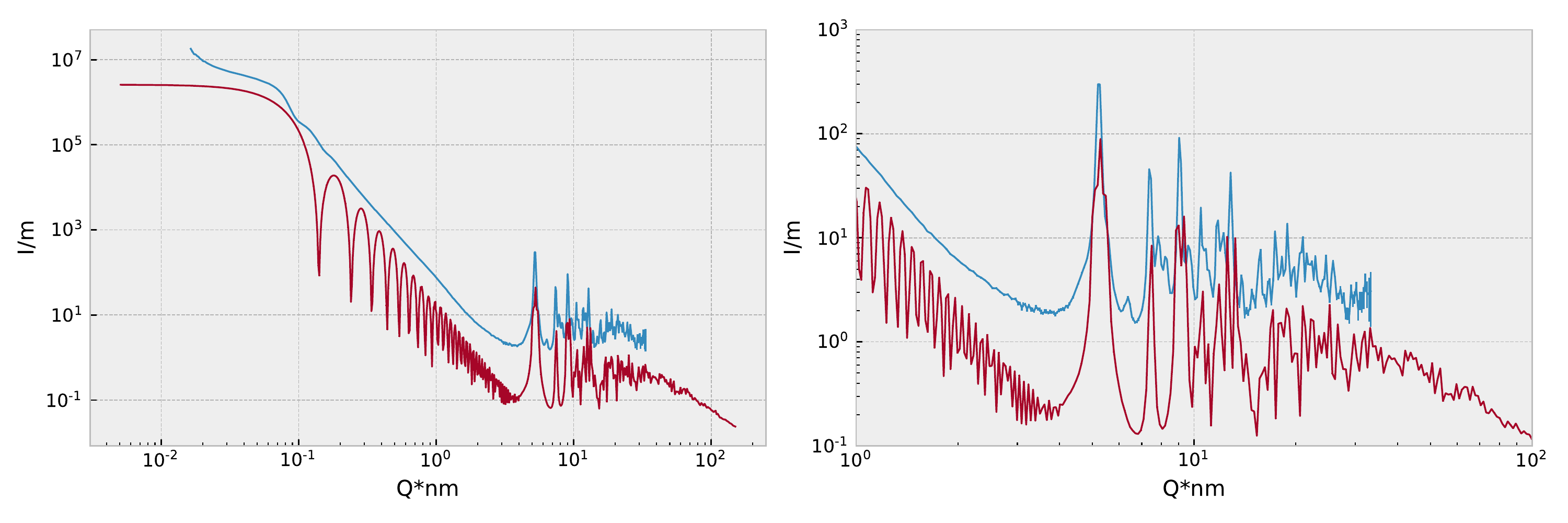}
\caption{Comparison between the merged simulated curve (red) and the measurement in our laboratory of ZIF-8-type MOF particles (blue). The left panel shows the full range, the right-hand side shows only the wide-angle and PDF range. } 
\label{fg:simvsmeas}
\end{figure}

The resulting simulated X-ray scattering curve, spanning the scattering vector range covering Small-angle to Wide-angle to PDF is shown in Figure \ref{fg:simvsmeas}. It is important to note that simulations of such box sizes are not possible in the standard approach, but with the separated approach they are, and take less space due to efficient intermediate and final storage. A MOUSE wide-range X-ray scattering measurement of one of the experimental, nominally 65\,nm MOF particles is added for comparison. The agreement over the entire range is remarkable, but does benefit from some notes. 

In the small-angle range, the difference between a single particle and an ensemble of polydisperse particles is evident. On top of that, there is additional scattering at the lowest Q from the state of the material (powder) during the measurements. The location of the bump maxima matches reasonably, but the question is raised on how to address polydispersity. From our experience with developing the McSAS software \cite{Bressler-2015}, we know it is possible to accurately describe even broad polydispersity with only 200 volume-weighted contributions. That would translate here to performing 200 simulations which, while large, is doable within reasonable time spans, especially if the computational hardware and code efficiency is improved. This would have the added benefit of exactly describing the peak broadening effects in the wide-angle range. 

Even with the single particle simulation, the wide angle range matches very well, both in terms of peak broadening as well as peak locations. While there is a discrepancy between one or two peaks, this is not unsurprising as 1) the sample may contain traces of contaminants, and 2) the simulation is restricted to the density generated from an idealized lattice and may therefore not contain reflections from imperfect crystals. Looking towards even wider angles, the PDF range shows a gradual drop in intensity, which matches the experimental data as far as they have been collected.

With the diffraction information now available relative to the small-angle signal, a new pathway for determining crystallinity may have opened up. As the simulation assumes perfect crystallinity, the integral intensity under the simulated diffraction peak should correspond to 100\% particle crystallinity, and can be assessed in absolute intensity units, or relative to the small-angle scattering signal. Armed with this knowledge, actual integral peak intensities of practical measurements may be compared to the simulated purely crystalline patterns and should enable the extraction of an actual degree of crystallinity. 

Overall, this demonstration provides sufficient evidence that high-resolution, wide-range simulations are possible for X-ray scattering. The previous limitations of resolution have been overcome, provided enough disk space and computational capacity is available. The bottleneck has now been shifted to real space, where the limits are on the generation of electron density maps that are a sufficiently accurate reflection of the structure in reality. Toolsets and libraries to address these are available, but likely need adapting to be able to generate high-resolution matrices in chunks. We expect to see this approach used in the near future by ourselves as well as others for a wide range of scattering-related investigations.

\section{Future steps}

While the concept has been demonstrated, there is much to gain by having dedicated HPC (high performance computing) programming experts improve the underlying code. Dramatic speed increases may be achievable through: 
  1. clever use of memory and disk I/O, e.g. by using the "zarr" library instead of HDF5 as "zarr" supports simultaneous writing, 
  2. improved parallelization and separation between multiple threads and multiple processes, potentially with automatic tuning. 
  3. better profiling to identify and alleviate processing bottlenecks, and 
  4. optional offloading to GPU hardware for Faster Fourier transforms. 

Undoubtedly, the seasoned programmer will be able to identify several more improvements, and it is expected that at least a factor 10 of speed improvements can be uncovered given enough time. 

Secondly, the electron density generation can be worked on to improve the realism of the obtained particle density. This could include polycrystalline and amorphous particle grains as well as better truncation at the particle edges, the scipy "ndimage" library might be a good starting point for such constructions, for example using biased erosion on the particle edge to improve its boundary [https://docs.scipy.org/doc/scipy/tutorial/ndimage.html]. Alternatively, other packages such as \verb|PoreSpy| \cite{Gostick-2019} or \verb|XRD_simulator| \cite{Henningsson-2023} be exploited to generate such realistic structures, which could be elaborated by filling in the different phases with a crystalline electron density map. Additionally, the surrounding matrix could be set to a more realistic density such as that of water with local density fluctuations (which we can find through the isothermal compressibility). 

Lastly, the benefit of merging several simulations at different length scales may be exploited to calculate the same or wider range faster with reduced resources. Additionally, a multiphase system, such as one consisting of several different types of particles supported by a porous carbon matrix can be simulated by combining the individual components at different scales.

\section{Summary}

With the memory limitations circumvented, and by combining multiscale simulations, the Fourier Transform approach offers a viable pathway to simulations of wide-range scattering patterns. These simulated datasets can be compared directly to real scattering data, be used for testing of analytical models, or be used to test the feasibility of experiments by comparing signal strength to background levels. Apart from time and disk space, the remaining limitations to this approach are the implementability of the electron density models, and the imagination of the researcher.

\section{Acknowledgments}
BRP would like to thank Dr. Steen L. Hansen for his offhand hints many years ago that the FFT could be split up. Thanks to Dr. Thilo Muth for making Sofya Laskina available to explore this problem. 

\section{Bibliography}

\bibliographystyle{vancouver}
\bibliography{main}

\end{document}